\def\pd#1#2{ { \partial #1 \over \partial {#2} } }
\def\pdd#1#2{ { \partial^2 #1 \over \partial {#2}^2 } }
\newcommand{\pddd}[3]{ { \partial^2 {#1} \over \partial {#2} \partial {#3} } }
\def\ImI{ {\rm I} }

\newcommand{\llabel}[1]{\label{#1}}                 

\documentclass[11pt]{article}      
\usepackage{amstext,amsmath}

\title{\bf Gauge stability of 3+1 formulations \\of General Relativity}  

\author{A.M Khokhlov\footnote{Laboratory for Computational Physics, Code 6404, Naval Research Laboratory, Washington, DC 20375}
              and  
        I.D. Novikov\footnote{Theoretical Astrophysics Center, Juliane Maries vej 30, DK-2100 Copenhagen, Denmark}
                    \footnote{Copenhagen University Observatory, Juliane Maries vej 30, DK-2100 Copenhagen, Denmark}
                    \footnote{Astro Space center of the P.N. Lebedev Physical Institute, Profsoyouznaja 84/32, Moscow 118710, Russia}
                    \footnote{NORDITA, Blegdamsvej 17, DK-2100 Copenhagen, Denmark} 
       }

\begin{document}

\maketitle                   

\large

\begin{abstract}
We present a general approach to the analysis of gauge stability of 3+1 formulations of General Relativity (GR). 
Evolution of  coordinate perturbations   and
the corresponding perturbations of lapse and shift can be described by a system of 
eight  quasi-linear partial differential equations. 
Stability with
respect to gauge perturbations depends on a choice of gauge and a background metric, but it does not depend on a
particular form of a 3+1 system if its constrained solutions  are equivalent to those of the Einstein equations. 
Stability of a number of known gauges is investigated in  
 the limit of short-wavelength perturbations. 
All fixed gauges except a synchronous gauge are found to be ill-posed. A 
maximal slicing  gauge  and its parabolic extension are shown to be ill-posed as well. A 
necessary condition is derived for well-posedness of  metric-dependent algebraic gauges. 
Well-posed metric-dependent gauges are found, however, to be
generally unstable.  Both instability and ill-posedness
are associated  with perturbations of physical accelerations of reference frames.    
\end{abstract}

\clearpage

\numberwithin{equation}{section}

\section{Introduction}

Physical analysis of many problems of general relativity (GR) requires a solution of  a full set of the non-linear
Einstein equations. Due to a complexity of these equations,  in most cases this can be done only  numerically.
This paper is concerned with gauge stability of  3+1 approaches to numerical integration of the Einstein equations. 
3+1 means here any approach in which the equations are split on constraint and  evolution parts. Constraints are satisfied
on an initial three-dimensional space-like hypersurface. Initial data is then  evolved in time by solving a 
Cauchy problem. 

The original ADM 3+1 formulation uses a three-dimensional metric $\gamma_{ij}$  and an extrinsic curvature
$K_{ij}$
 as unknown functions \cite{ADM}.
In other 3+1 formulations,  such as 
hyperbolic, conformal formulations and their combinations, 
 the ADM equations are extended by
introducing additional variables such as spatial derivatives of $\gamma_{ij}$, traces of $\gamma_{ij}$ and $K_{ij}$, conformal
factors,  by forming new combinations of these variables, or by modifying equations with the help of
constraints [2-18].
\nocite{BONA-89,ABRAHAMS-95,FRITTELLI-96,BONA-97,BONA-97-1,ANDERSON-98,ARBONA-99,ALCUBIERRE-99,BAUMGARTE-99,BRODBECK-99,LAGUNA-99,SCHEEL-99,
        SHIBATA-99,SHIBATA-99-2,ALCUBIERRE-00-1, SHINKAI-00,KELLY-01,KIDDER-01} 
These modifications change the nature of the equations so that they can  become
more stable, and the integration can be prolonged [19-23].
\nocite{ANNINOS-95-2,BONA-98,SEIDEL-98,SEIDEL-99,ALCUBIERRE-01-2}
 Nonetheless, calculations (of black hole collisions, in particular) often suffer from instabilities.  A
stable, accurate, long-term integration of the Einstein equations remains as an outstanding problem of numerical GR. 

Difficulties with time-integration  can be 
 numerical and analytical. Numerical difficulties  arise when an unstable finite-difference scheme is
used to integrate a particular 3+1 set of  equations. Constructing a stable numerical scheme 
for a chosen 3+1 set of equations should not be a problem if the  equations are well-posed and stable.  Numerical stability
will not be discussed in this paper ( see, e.g.,  \cite{ALCUBIERRE-00-1,MILLER-00, TEUKOLSKY-00} for numerical aspects). We believe
that  analytical difficulties  arising from the nature of the Einstein equations themselves present a much more serious and still
unsolved problem.  

First of all,   a 3+1 set of GR
equations must be well-posed  \cite{CH-II,WILLIAMS-80}. 
It is impossible to guarantee  for an ill-posed system a convergence
of solutions when numerical resolution is  increased. It is known, for example, 
that a harmonic gauge leads to a  symmetric hyperbolic,  well-posed  system \cite{FISHER-72,WALD}. 
Some gauges  lead to
 ill-posed  systems (see below). The property of well-posedness is local and time-dependent.
  Using a  gauge that
maintain well-posedness everywhere  and at all times is crucial for a 
long-term stable integration. 

Maintaining well-posedness, however, is not enough. 
A well-posed non-linear
system  may  still have rapidly growing, diverging, or singular solutions so that numerical integration will become
problematic. Here we call  this  an analytical instability.
Analytical instabilities can be separated on three types. First, the instability may be related to a physical nature of
 space-time. For example, one can encounter a true singularity where curvature invariants become infinite.  Second, the
instability may be related 
 to violation of constraints during time-integration (constraint instability).  These may be the energy and momentum
constraints, as well as  additional constraints arising from introduction of additional variables into a
system.
 If  constraints are satisfied initially, they are automatically
satisfied  at later times. However, a Cauchy problem  possesses a 
 much broader class
of solutions than  constraint-satisfying solutions. Consequently, a small initial perturbation may lead to a rapid
deviation from a constrained solution during integration. 
The third type 
 is the gauge instability. In a 3+1 approach, a coordinate system is  ``constructed" during time
integration according to a pre-determined choice of gauge and initial conditions. Initially small coordinate perturbations 
 may lead to a  divergence of  coordinate systems  as time progresses. For a long-term integration to be successful, a gauge 
must be  both well-posed and stable.

Little is known about  stability properties of  3+1 systems of GR equations in general. The analysis
is complicated because many different modes are present simultaneously, stability is determined by complex
non-linear terms in the Einstein equations, and it depends on a particular solution and on location in spacetime.   
Investigation of well-posedness is somewhat easier since it requires  analyzing only a  principal part of the system. Several
classes of  hyperbolic, well-posed 3+1 formulations have been constructed (e.g.,  
\cite{BONA-97, BAUMGARTE-99,SCHEEL-99,KELLY-01,KIDDER-01}), but  
 numerical experiments show that these systems are often unstable. There are indications that, at least
partially, the instability is related to constraint violating modes (e.g.,  \cite{SCHEEL-99, KIDDER-01}). 

It is long
known that a  synchronous gauge 
 is prone to the formation of coordinate singularities (or caustics) \cite{LL,LIFSHITS-61}. 
 Consequences of this  for a numerical integration
were  discussed most recently in
\cite{ALCUBIERRE-00-2}. An attempt to characterize the development of coordinate pathologies (shocks) in 
hyperbolic 3+1 formulations has been made in \cite{ALCUBIERRE-98}.  Perturbations of coordinates
have been  separated, in linear approximation,  and studied for a synchronous gauge (see Problem 3 in
Paragraph 95 of \cite{LL} and also \cite{ALCUBIERRE-00-2}). To our knowledge,   instability  of gauges
 other than a synchronous gauge  have  not been analyzed.

In this paper we  develop a general approach to the analysis of gauge instabilities.  
In Section 2 we  derive a system of eight quasi-linear partial 
differential equations that describe coordinate perturbations for
arbitrary gauges and in arbitrary 3+1 system.  We
illustrate our approach by
 investigating stability of various gauges in Section 3.  A physical meaning of  gauge instabilities  is
discussed in Section 4. Our conclusions are given in Section 5. 


\section{Equations of gauge perturbations}

Consider a space-time described in a certain coordinate system $x^a$ by a four-dimensional metric
$g_{ab}$. In what follows we use letters $a-h$ to denote four-dimensional indices  $0,1,2, 3$, 
and letters $i-m$ to denote  three-dimensional spatial indices $1,2,3$;  time $t= x^0$.
  We write an interval as \cite{MTW}
\begin{equation}\llabel{eq:interval}
           ds^2 = g_{ab}dx^adx^b = -(\alpha^2 - \beta_i\beta^i)dt^2 + 2\beta_i dt dx^i + \gamma_{ij}dx^idx^j  
\end{equation}
where $\alpha$ is a lapse, $\beta_i$ is a shift, $\beta^i=\gamma^{ij}\beta_j$,
$\gamma^{im}\gamma_{mk}=\delta^i_k$,  $\gamma_{ij}$ is a metric on a three-dimensional space-like hypersurface, and the
4-metric is
\begin{equation}\llabel{eq:metric}
              g_{ab}~=~ \begin{pmatrix} 
                               -(\alpha^2 +\beta_i\beta^i)         &{\beta_i }      \cr
                               {\beta_i}                         &\gamma_{ij} \cr
                        \end{pmatrix}~,~~~g^{ab}~=~     \begin{pmatrix} 
                               -{1\over\alpha^2}         &{\beta^i\over\alpha^2 }      \cr
                               {\beta^i\over\alpha^2}    &\gamma^{ij}-{\beta^i\beta^j\over\alpha^2} \cr
                \end{pmatrix}~,~~~g_{ab} g^{bc} = \delta_a^c~.
\end{equation}

We want to separate, in a linear approximation, the behavior of coordinates from physical evolution of a space-time.
 Under an infinitesimal coordinate (gauge) transformation $\psi^a(x^0,x^1,x^2,x^3)$  of a
four-dimensional coordinate system, 
\begin{equation}\llabel{eq:pertsys}
x^a \rightarrow x^a + \psi^a~,  
\end{equation}
where $\psi^a$ are sufficiently smooth infinitesimal functions, the metric transforms as \cite{LL}
\begin{equation}\llabel{eq:pertmet}
                 g_{ab} \rightarrow g_{ab} + \delta g_{ab}~, 
\end{equation}
where 
\begin{equation}\llabel{eq:varmetric}
\delta g_{ab} = -(\nabla_b\psi_a + \nabla_a\psi_b)  
\end{equation}
are the deviations of the metric, 
and  $\nabla$ denotes a covariant derivative.

For a metric \eqref{eq:interval}, the deviations $\delta g_{ab}$ can also be related to 
corresponding deviations of lapse, shift, and three-dimensional metric,
\begin{equation}
U\equiv\delta\alpha~, ~~ V_i \equiv \delta\beta_i~,~~  W_{ij} \equiv \delta\gamma_{ij}
\end{equation} 
 as
\begin{equation}\llabel{eq:varmetric1}
\begin{split}
\delta g_{00} & = -2\alpha U + 2\gamma^{ij} \beta_i V_j - \gamma^{ik}\beta_i\beta_j W_{jk} \cr
\delta g_{0i} & = V_i \cr
\delta g_{ij} & = W_{ij} \cr
\end{split}
\end{equation}
Substitution of \eqref{eq:varmetric1} into \eqref{eq:varmetric} gives us $U$, $V_i$ and $W_{ij}$  
in terms of $\psi_a$,
\begin{equation}\llabel{eq:U}
U = {1\over \alpha} \left( {\beta^i\beta^j W_{ij} \over2} - \beta^i V_i - \pd{\psi_0}{t} + \Gamma^c_{00}\psi_c \right)
~,                                                                                                        
\end{equation}
\begin{equation}\llabel{eq:Vi}
V_i     = -\left(\pd{\psi_i}{t} +\pd{\psi_0}{x^i}\right)  + 2\Gamma^c_{0i} \psi_c  ~,         
\end{equation}
\begin{equation}\llabel{eq:Wij}
W_{ij} = -\left( \pd{\psi_j}{x^i} +\pd{\psi_i}{x^j} \right) + 2\Gamma^c_{ij} \psi_c  ~,      
\end{equation}
where $\Gamma^c_{ab}$ are four-dimensional Cristoffel symbols.   They can be expressed   in terms of
$\alpha$,
$\beta_i$, and
$\gamma_{ij}$ using \eqref{eq:metric}.
Equations 
 \eqref{eq:U} and \eqref{eq:Vi} can also be rewritten as a system of  four quasi-linear partial differential equations for
$\psi_a$,
\begin{equation}\llabel{eq:gauge}
\begin{split}
&\pd{\psi_0}{t} = \alpha U - \beta^i V_i - \beta^i\beta^j\pd{\psi_i}{x^j} 
               + \left(\Gamma^c_{00} + \beta^i\beta^j \Gamma^c_{ij} \right) \psi_c   ~, \cr
&\pd{\psi_i}{t} = - V_i - \pd{\psi_0}{x^i} + 2 \Gamma^c_{0i} \psi_c~.  \cr
\end{split}   
\end{equation}
The deviations $\psi^a$, and the corresponding deviations  $U$, $V_i$ were arbitrary up to this point.
  
During the integration of a 3+1 system  in time, certain conditions are imposed on 
$\alpha$ and $\beta_i$ which specify  the choice of gauge. In general, a gauge  can be specified  by a set  of four
partial differential equations with respect to lapse, shift, three-dimensional metric, and  their partial derivatives,
\begin{equation}\llabel{eq:GaugeChoice}
F_a\left(x^b,\alpha,\pd{\alpha}{x^b},...,\beta_i,\pd{\beta_i}{x^b},...,\gamma_{ij}, \pd{\gamma_{ij}}{x^b},...\right) = 0~,
\end{equation}
where $...$ in \eqref{eq:GaugeChoice} indicate higher order derivatives of $\alpha$, $\beta_i$ and $\gamma_{ij}$. Equations
\eqref{eq:GaugeChoice} must be considered a part of a 3+1 system which is integrated in time along with the rest
of the equations for metric components, extrinsic curvature, etc.  By  varying \eqref{eq:GaugeChoice} with respect
to $\alpha$, $\beta_i$ and $\gamma_{ij}$ and making use of \eqref{eq:Wij} we obtain a system of four additional quasi-linear 
partial differential equations relating  $U$ and $V_i$ to coordinate perturbations $\psi^a$,
\begin{equation}\llabel{eq:CaseDifferential}
\begin{split}
&\pd{F_a}{\alpha} U + \pd{F_a}{\left(\pd{\alpha}{x^b}\right)}\pd{U}{x^b} +
\pd{F_a}{\left(\pddd{\alpha}{x^b}{x^c}\right)}\pddd{U}{x^b}{x^c} +...+ \cr
&\pd{F_a}{\beta_i} V_i +
\pd{F_a}{\pd{\beta_i}{x^b}}\pd{V_i}{x^b} + \pd{F_a}{\left(\pddd{\beta_i}{x^b}{x^c}\right)} \pddd{V_i}{x^b}{x^c} +...+ \cr
 &  2\,\pd{F^a}{\gamma_{ij}} \left( \Gamma^c_{ij} \psi_c
-\pd{\psi_j}{x_i}   \right) 
      + 2\, \pd{F^a}{  \left(\pd{\gamma_{ij}} {x^b}\right)  } {\partial\over\partial x^b}\left( \Gamma^c_{ij} \psi_c
-\pd{\psi_j}{x_i}  
               \right)+ ... = 0\cr
\end{split}
\end{equation}
Together, \eqref{eq:gauge} and \eqref{eq:CaseDifferential} form a system of eight  equations describing
 the evolution of coordinate perturbations and associated perturbations of lapse and shift in time.
Perturbations of a three-dimensional metric which correspond to evolving coordinate perturbations are given by \eqref{eq:Wij}.

At this point, we succeeded in separating the behavior
of gauge perturbations from other possible perturbations of the solutions of a 3+1 system. Gauge instabilities can be
studied by  investigating  a set of eight quasi-linear partial differential
equations
\eqref{eq:gauge} and \eqref{eq:CaseDifferential} for coordinate perturbations $\psi_a$ and associated perturbations of lapse and
shift, $U$ and $V_i$. During the derivation, we did not use any specific assumptions
about a 3+1 system. The only assumption implicit to the derivation was the equivalence of 
constrained solutions of a 3+1 system to these of the  Einstein equations. Therefore, 
\eqref{eq:gauge} and \eqref{eq:CaseDifferential} must be applicable to any such system.  In a
linear approximation, the  behavior of gauge modes of perturbations depends only on a choice of gauge and on an unperturbed
metric $g_{ab}$.

Since coefficients of \eqref{eq:gauge}, \eqref{eq:CaseDifferential} are  functions of $g_{ab}$ and thus of
$x^a$, exact solutions of \eqref{eq:gauge}, \eqref{eq:CaseDifferential} can  be found only in some  special cases.   A further 
simplification is possible in a high-frequency limit where we consider  coordinate perturbations  on  much shorter scales
compared to  those of a  base solution $g_{ab}$. Let us designate
\begin{equation}
\vec{ z}\,(x^a) = \{\psi^a,U,V_i \}^T
\end{equation}
 a combined vector of
unknowns  entering our gauge perturbation equations. We will consider in this paper the gauge
equations \eqref{eq:CaseDifferential} which involve $\gamma_{ij}$ and its first-order time derivatives (cases with higher order
derivatives can be treated similarly). Then the resulting gauge perturbation equations will contain only first-order time derivatives of
$\vec z$, but, of course, they can contain higher-order spatial derivatives of $\vec z$. We  write  \eqref{eq:gauge},
\eqref{eq:CaseDifferential} symbolically as
\begin{equation}\llabel{eq:eqpert}
\pd{ z_r}{t} = {^{(0)}{\cal M}_{rs}} z_s + {^{(1)}{\cal M}^i_{rs}}\pd{z_s}{x^i}  + {^{(2)}{\cal M}^{ij}_{rs}}\pddd{z_s}{x^i}{x^j} + ...,
\end{equation}
where  ${^{(k)}{\cal M}}$, $k=0,1,2, ...$  depend on an unperturbed  solution.
In the limit of short-wavelength perturbations, 
we look for solutions of \eqref{eq:eqpert} in the form
\begin{equation}\llabel{eq:pert}
\vec z = \vec \zeta(t) \exp  ( - \ImI q e_k x^k )~,
\end{equation}
where $\ImI=\sqrt{-1}$, $q_i = q e_i$ is a wavevector, $e_i$ is a unit vector, and $q$ is an
absolute value of
$q_i$. 
 Substituting \eqref{eq:pert} into \eqref{eq:eqpert}  gives 
an ordinary differential equation for $\vec \zeta$ (we retain only the time-dependence of ${^{(k)}{\cal M}}$),
\begin{equation}\llabel{eq:eqpert1}
{d \vec \zeta\over dt } =  \hat M(t) \, \zeta ~,
\end{equation}
with
\begin{equation}\llabel{eq:eqpert2}
M_{rs}(t,e_i,q)  = {^{(0)}{\cal M}_{rs}(t)} + {^{(1)}{\cal M}^i_{rs}(t)}  \ImI e_i q + {^{(2)}{\cal M}^{ij}_{rs}(t)} e_i e_j  q^2 + ...
\end{equation}
When the time-dependence of $\hat M$ can be neglected, the solutions become
\begin{equation}\llabel{eq:Harmpert}
\vec{ z}\,(x^a) \propto \exp  ( \omega_s t - \ImI q e_k x^k ) ~,      
\end{equation}
where wavenumbers  $\omega_s(q,e_i)$ are determined by the dispersion relation 
\begin{equation}\llabel{eq:eigenval}
det \left( \hat M - \omega \hat N \right) = 0~,
\end{equation}
and $\hat N$ is a unit matrix.
If a timescale of the variation of coefficients in $\hat M$ becomes comparable to $\vert\omega\vert^{-1}$, the behavior of perturbation
with time will not be exponential any more. However, a time derivative of $\vec z$ at $t=0$ will still be determined
by the  corresponding value of
$\omega$.
 In particular, the growth rate of the  perturbation amplitudes  will be given by
$Re(\omega_s)$.

An information about $Re(\omega_s)$
  can be used, first of all, to probe ill-posedness of a gauge.
If  ${\rm Re}(w) \rightarrow\infty$  when
$ q \rightarrow \infty$ for at least one  $\omega_n$ and at least in one direction  $e_k$,  it would be possible to
construct a harmonic perturbation  such that at $t=0$ both the perturbation and any finite number of its derivatives 
 will be  less than any predetermined small number $\epsilon << 1$  but will become greater than any large predetermined number
$A >> 1$ after a finite period of time.  This will mean, of course,  that the gauge is ill-posed. 
 If, on the other hand, the
gauge is well-posed\footnote{The absence of short-wave harmonic solutions with arbitrary
large real increments is a necessary condition for well-posedness. Sufficient condition would require proving a continuous
dependence of solutions on initial conditions for the entire 3+1 system.}, than
${\rm Re}(\omega_s)$  will determine the rate of growth of instabilities with
time.


\section{Gauge stability}

In what follows, it will be convenient to distinguish between the three types of gauges.
A gauge is called ``fixed" if none of $F_a$ in \eqref{eq:GaugeChoice}  involve  $\gamma_{ij}$, and $F_a(x^b,\alpha,\beta_i) =0$.
We assume that \eqref{eq:GaugeChoice} can  be inverted and $\alpha$,
$\beta_i$ can be  expressed  explicitly as functions of four-coordinates, 
\begin{equation}\llabel{eq:fixed}
 \alpha=\alpha(x^a)~,~~~\beta_i=\beta_i(x^a)~.
\end{equation}
A fixed gauge does not change when coordinates are perturbed, $\pd{\alpha}{\psi_a} = \pd{\beta_i}{\psi_a}=0$, and, as a
result, 
$U=V_i=0$. Equations \eqref{eq:gauge} with  $U=V_i=0$ describe the evolution of coordinate perturbations for  fixed
gauges.  

A gauge is called ``algebraic" or ``local" if it can be expressed as a  function of local values of $\gamma_{ij}$
and its derivatives,  $F_a\left(x^b,\alpha,\beta_i,\gamma_{ij}, \pd{\gamma_{ij}}{x^b},...\right) = 0$, or
\begin{equation}\llabel{eq:AlgebraicGaugeChoice}
 \alpha=\alpha(x^a,\gamma_{ij},
\pd{\gamma_{ij}}{x^b},...)~,~~~\beta_i=\beta_i(x^a,\gamma_{ij},
\pd{\gamma_{ij}}{x^b},...)~.
\end{equation}
 Fixed gauges are, of course, a particular case of algebraic gauges.
Expressions for $U$ and $V_i$ for algebraic gauges reduce  to 
\begin{equation}\llabel{eq:CaseAlgebraic}
\begin{split}
U  = &2\,\pd{\alpha}{\gamma_{ij}} \left(  \Gamma^c_{ij} \psi_c - \pd{\psi_j}{x_i}     \right) + 
     2\,\pd{\alpha}{  \left(\pd{\gamma_{ij}} {x^a}\right)  } 
            {\partial\over\partial x^a}\left( \Gamma^c_{ij} \psi_c  - \pd{\psi_j}{x_i} \right) + ... \cr
V_k  = &2\,\pd{\beta_k}{\gamma_{ij}} \left(  \Gamma^c_{ij} \psi_c  - \pd{\psi_j}{x_i}  \right) + 
     2\,\pd{\beta_k}{  \left(\pd{\gamma_{ij}} {x^a}\right)  } 
            {\partial\over\partial x^a}\left( \Gamma^c_{ij} \psi_c - \pd{\psi_j}{x_i}  \right) + ... \cr 
\end{split}
\end{equation}
Finally,  a gauge  is called ``differential" or ``non-local" if it cannot be reduced from \eqref{eq:GaugeChoice} to an 
algebraic form. 
An algebraic gauge can always be
expressed in a differential form by simply  differentiating
\eqref{eq:AlgebraicGaugeChoice}.  Values of $\alpha$ and $\beta_i$ for a differential gauge can
only be expressed as  a non-local functional of $\gamma_{ij}$.

\subsection{Fixed gauges}

To illustrate the approach outlined above, we begin with gauges \eqref{eq:fixed} which are functions of spatial coordinates and
time only.
A synchronous gauge $\alpha=\alpha(t)$, $\beta_i=0$ is an important member of this family. We will show below that it is the
only  well-posed fixed gauge. All other gauges \eqref{eq:fixed} are ill posed.

For a fixed gauge, \eqref{eq:gauge} become a first-order linear system of partial differential
equations
\begin{equation}\llabel{eq:fixedgauge}
\begin{split}
&\pd{\psi_0}{t} =  - D^{ij} \pd{\psi_i}{x^j} + C^a  \psi_a   ~, \cr
&\pd{\psi_i}{t} =  - \pd{\psi_0}{x^i} + E^a_i \psi_a~.     \cr
\end{split}                                                                                 
\end{equation}
where
\begin{equation}\llabel{eq:Cafix}
C^a = \Gamma^a_{00} + D^{ij} \Gamma^a_{ij}~,  
\end{equation}
\begin{equation}\label{eq:Dijfix}
D^{ij} = \beta^i\beta^j ~, 
\end{equation}
and 
\begin{equation}\llabel{eq:fixEak}
E^a_k = 2 \Gamma^a_{0k}~. 
\end{equation}

For  synchronous gauges  \eqref{eq:fixedgauge}  becomes 
\begin{equation}\llabel{eq:syncgauge}
\begin{split}
\pd{\psi_0}{t} & = {1\over \alpha}\pd{\alpha}{t} \psi_0\cr
\pd{\psi_i}{t} & = -\pd{\psi_0}{x^i} + \gamma^{km}\pd{\gamma_{mi}}{t} \psi_k ~.\cr
\end{split}
\end{equation}
A general solution of \eqref{eq:syncgauge} is
\begin{equation}\llabel{eq:syncgauge1}
\begin{split}
&\psi_0 = \alpha(t) f_0~,  \\
&{\psi_i} = \gamma_{ij} \left( f^i - \pd{f_0}{x^k} \int^t_0 \alpha \gamma^{jk} dt  \right) ~,\\
\end{split}
\end{equation}
where $f_0(x^k),f^i(x^k)$ are smooth arbitrary function of spatial coordinates (see Problem 3 in Paragraph 97  of \cite{LL}).
 According to \eqref{eq:syncgauge1}, there
is   a
 continuous dependence of solutions on initial conditions in the
class of functions with continuous first derivatives. Thus a synchronous gauge 
 is  well posed.

Next, consider  solutions of \eqref{eq:fixedgauge} in a short-wavelength limit.  The matrix $\hat M$ in \eqref{eq:eqpert2} becomes
\begin{equation}\llabel{eq:syncgauge2}
\hat M = \hat M_0 + \ImI q \hat M_1
\end{equation}
with
\begin{equation}
 M_0 = \left( \matrix    C^0   &C^1  &C^2  &C^3 \\
                         E^0_1 &E^1_1 &E^2_1  &E^3_1 \\
                         E^0_2 &E^1_2 &E^2_2 &E^3_2 \\
                         E^0_3 &E^1_3 &E^2_3 &E^3_3 \\
                        \endmatrix
        \right) ~, ~~~~
\hat M_1 = \left( \matrix
                    0    &D^{1k}e_k & D^{2k}e_k & D^{3k}e_k\\
                     e_1 &0 &0 &0\\
                     e_2 & 0 & 0 &0 \\
                    e_3 & 0 & 0 &0\\
                  \endmatrix
           \right)~,
\end{equation}
and \eqref{eq:eigenval} becomes a quartic  equation for $\omega$,
\begin{equation}\llabel{eq:fixdisp1}
\omega^4 + d_3 \omega^3 + d_2 \omega^2 + d_1 \omega + d_0 = 0~, 
\end{equation}
with coefficients being polynomial functions of $q$,
\begin{equation}\llabel{eq:dixdi}
\begin{split}
d_0 &= d_{0,0} +  d_{0,1}~ q  + d_{0,2}~q^2 ~, \cr
d_1 &= d_{1,0} +  d_{1,1}~ q  + d_{1,2}~q^2 ~, \cr
d_2 &= d_{2,0} +  d_{2,1}~ q  + d_{2,2}~q^2 ~, \cr
d_3 &= d_{3,0} ~, \cr
\end{split}
\end{equation}
where $d_{i,j}$ are functions of $\gamma_{ij}$, $\alpha$, $\beta_i$, and   $e_i$. In what follows, we  need
expressions for 
\begin{equation}\llabel{eq:fixdij}
\begin{split}
&d_{0,2} = \left( E^i_m E^j_i - E^n_n E^j_m \right) D^{km} e_k  e_j  
          + {1\over 2} \left( E^i_i E^j_j  -  E^i_j E^j_i \right)  d_{2,2} ~, \cr
&d_{1,2} =  E^i_k D^{kj}  e_i e_j - E^k_k \, d_{2,2}  ~, \cr
&d_{2,1} = -I ( C^i +  E^0_k D^{ki} ) e_i  ~,\cr
&d_{2,2} = D^{ij} e_i e_j~, ~\text{and}\cr
& d_{3,0} = - C^0 - E^k_k~. \cr
\end{split}  
\end{equation}
From \eqref{eq:fixdisp1} and \eqref{eq:fixdij} it is clear that the asymptotic behavior of roots of \eqref{eq:fixdisp1} with
$q\rightarrow \infty$ must be $\omega = \sum \omega_k q^{k/m}$, with
$n$, $m$  integer. Values of $m,n$ and coefficients $w_k$ can be determined by
substituting a power series expressions for $\omega$ into \eqref{eq:fixdisp1} and  requiring that
terms with same powers of
$q$ cancel out. The result depends on whether some of $d_{i,j}$ are  zero or not. 

 If $d_{2,2}
\neq 0$, the asymptotic behavior of the roots is  
\begin{equation}\llabel{eq:fixasymp1}
\omega = \omega_1 q + \omega_0 + O\left({1\over q}\right) ~,
\end{equation}
 with
\begin{equation}\llabel{eq:fixw1}
(\omega_1)_{1,2,3,4} = \{0, 0,  \pm \ImI \sqrt{ d_{2,2} }\}  ~, 
\end{equation}
\begin{equation}\llabel{eq:fixw0}
(\omega_0)_{1,2,3,4} = \left\{ -{d_{1,2} \over 2 d_{2,2}} \pm \sqrt{ \left(d_{1,2}\over 2 d_{2,2}\right)^2 - 
{d_{0,2}\over d_{2,2}}}~,~~
                                -d_{3,0} \pm \ImI{  d_{2,1}\over \sqrt{d_{2,2}}} +{d_{1,2} \over d_{2,2}} 
                        \right\}~.
\end{equation}
 If $d_{2,2} =0$ but $d_{1,2} \neq 0$,  the asymptotic behavior changes to 
\begin{equation}\llabel{eq:fixasymp2}
\omega = \omega_2 q^{2/3} + \omega_1 q^{1/3} + \omega_0 + O\left({1\over q^{1/3}}\right) ~,
\end{equation}  
where
\begin{equation}\llabel{eq:fixw2}
\begin{split}
(\omega_2)_{1,2,3,4} = {\rm Sign}(d_{1,2}) \cdot \vert d_{1,2}\vert^{1/3} 
                               \cdot \{\, 0, \, 1, \, -{1\over 2} \pm \ImI {\sqrt{3}\over 2}\,\}  
\end{split}    
\end{equation}
If both $d_{2,2}=d_{1,2}=0$, the asymptotic behavior changes again, this time  to 
\begin{equation}\llabel{eq:fixasymp3}
\omega = \omega_1 q^{1/2}  + \omega_0 + O\left({1\over q^{1/2}}\right) ~,
\end{equation}
with 
\begin{equation}\llabel{eq:fixw3}
(\omega_1)_{1,2,3,4} =         \pm   \left( - d_{2,1} \pm \sqrt{ -4 d_{0,2} +( d_{2,1})^2 }  \over 2  \right)^{1/2} 
\end{equation}
 The asymptotic behavior is 
$\omega
\sim O(1)$ if all three coefficients $d_{2,2}=d_{1,2}=d_{2,1}=0$. In particular, for a synchronous gauge $\alpha=1$, $\beta_i=0$, the
increments are determined by $det(\gamma^{km} \pd{\gamma_{mi}}{t} -  \omega \delta^m_i) =0$.

For fixed algebraic gauges with zero shift  we have $D^{ij} = 0$,
$d_{2,2}=d_{1,2}=0$, and from \eqref{eq:fixw3} the asymptotic behavior  is 
\begin{equation}\llabel{eq:fixzero}
\omega_{1,2}  = \, \pm \, q^{1/2} \, (1 - \ImI {C^ie_i\over \vert C^i e_i  \vert}  ) \sqrt{ \vert C^i e_i  \vert \over 2} + O(1)
 ~,~~\omega_{3,4}=O(1)~.
\end{equation}
$C_i$ in  \eqref{eq:Cafix} reduces  for  $\beta_i=0$  to
\begin{equation}\llabel{eq:fixCzero}
C^i = \Gamma^i_{00} = - \alpha \gamma^{ik} \pd{\alpha}{x^k}~.
\end{equation}
 Since
$e_i$ is a unit but otherwise  arbitrary vector, there always  will be harmonic solutions with $Re(\omega) \sim q^{1/2}$
and the gauge will be ill-posed unless lapse is spatially constant, $\alpha =\alpha(t)$.

Consider now gauges with $\beta_i\neq 0$. According to \eqref{eq:fixw1}, $Re(w) \sim O(1) $  for all wavevectors not
orthogonal to shift, $\beta^i e_i \neq 0$, so that these perturbations will not cause ill-posedness.  However, it can be seen
from
\eqref{eq:fixw0}   that for  small $\beta_i << 1$ the increment $Re(\omega) \simeq \pm{\beta^{-1} 
C^ie_i}$ can become arbitrary large.  Thus, a mode of perturbation  unstable for $\beta_i=0$
cannot be eliminated by applying a small shift in the direction of its propagation.    

For
wavevectors which are orthogonal to shift,
$\beta^i e_i=0$, we find 
$d_{1,2} = (  E^i_k D^{kj} - E^k_k D^{ij} ) e_i e_j  = 0 $, and  the asymptotic behavior
is again given by \eqref{eq:fixzero} but now with $C^i$ determined by the general formula \eqref{eq:Cafix}. If
$C^i\beta_i \neq 0$, all orthogonal harmonic solutions will have  $Re(\omega)\sim q^{1/2}$, and the gauge will be
ill-posed. We now show that this is always the case, that is, $C^i$ and $\beta_i$ cannot be co-linear.   Expressing 
$\Gamma^a_{bc}$ in terms of three-dimensional quantities, we find that $C^i$ can be written as
\begin{equation}
C^i =  \gamma^{ik}\left( \pd{\beta_k}{t} + \pd{(\beta^m\beta_m - \alpha^2)}{x^k} \right) + f(x^a) \beta^i ~. 
\end{equation}
Thus, co-linearity of $C^i$ and $\beta_i$ requires that $\beta_i$  must satisfy a system of partial
differential equations
\begin{equation}\llabel{eq:fixab}
\pd{\beta_k}{t} - \pd{(\alpha^2 - \gamma^{mn}\beta_n\beta_m )}{x^k} = 0~. 
\end{equation}
These equations depend on $\gamma^{nm}$ and, thus, $\beta_i$  must depend on $\gamma^{nm}$ as well. This
 contradicts, however, to our initial assumption \eqref{eq:fixed} that $\alpha$ and $\beta_i$ are explicit functions of spatial
coordinates and do not change when $\gamma^{nm}$ are perturbed. 
 It is easy to see that the only solution of \eqref{eq:fixab} independent of $\gamma^{mn}$ is $\beta_i=0$, and that this
solution is possible if  $\alpha = \alpha(t)$. We finally conclude that, unless shift is zero and lapse
is spatially-constant, there always will be initially arbitrary small harmonic
solutions of (3.11) which can be made arbitrary large after a finite period of time by an appropriate choice of $q_i$.
That is, among  fixed algebraic gauges \eqref{eq:fixed} only  synchronous gauges $\beta_i=0$, $\alpha =
\alpha(t)$ are well-posed. All other gauges \eqref{eq:fixed} are ill-posed\footnote{In a one-dimensional  case where
both a metric and its  perturbation are dependent on one spatial coordinate $x=x^1$, we have $\vec{e} = \{1,0,0\}$,  and a
non-zero shift along the $x$-coordinate leads to $\beta_i e_i \neq 0$ and to a well posed gauge. This however, is not a
contradiction  since  coordinate perturbations orthogonal to gauge were not allowed.}. 


\subsection{Algebraic gauges}

As a next example, consider gauges with metric-dependent lapse and fixed shift,
\begin{equation}\llabel{eq:metdep}
\alpha = \alpha(x^a,\gamma_{ij})~,~~\beta_i = \beta_i(x^a)       
\end{equation}
From \eqref{eq:CaseAlgebraic} we  obtain 
\begin{equation}\llabel{eq:metdepU}
U  = 2 \pd{\alpha}{\gamma_{ij}} \left( \Gamma^c_{ij} \psi_c - \pd{\psi_i}{x^j}\right)~, ~~~V_i=0~,
\end{equation}
and the coordinate perturbation equations \eqref{eq:gauge} become
\begin{equation}\llabel{eq:metdepgauge}
\begin{split}
\pd{\psi_0}{t} & = - \left( \beta^i\beta^j + \pd{\alpha^2}{\gamma_{ij}} \right)   \pd{\psi_j}{x_i}  
               + \left(\Gamma^a_{00} + \left( \beta^i\beta^j + \pd{\alpha^2}{\gamma_{ij}} \right)   \Gamma^a_{ij} \right)
\psi_a   \cr
\pd{\psi_i}{t} & =  - \pd{\psi_0}{x^i} + 2 \Gamma^a_{0i} \psi_a~.  \cr
\end{split}
\end{equation}
Comparing gauge perturbation equations \eqref{eq:metdepgauge} and \eqref{eq:fixedgauge} we observe that 
\eqref{eq:metdepgauge}  becomes identical to
\eqref{eq:fixedgauge} if  the definition of
$D^{ij}$ is changed from \eqref{eq:Dijfix} to 
\begin{equation}\llabel{eq:Dijmetdep}
D^{ij} = \beta^i\beta^j + \pd{\alpha^2}{\gamma_{ij}} ~.
\end{equation}
In what follows, we assume that $\pd{\alpha^2}{\gamma_{ij}}$ and , thus, $D^{ij}$ are symmetric. 
We obtain a  dispersion relation similar to \eqref{eq:fixdisp1}, and,
after  taking a limit of
$q\rightarrow\infty$,  get an asymptotic behavior of wavenumbers
$\omega$ which dependence on coefficients $d_{i,j}$ is similar to that described by equations
(\ref{eq:fixasymp1}-\ref{eq:fixw3}).
Now, however, $D^{ij} \neq \beta^i\beta^j$ and thus $d_{2,2} = D^{ij}e_ie_j$ may be both positive, zero, and negative.
For $d_{2,2} < 0$ we have two roots with $Re(\omega) \sim q$, and the gauge is ill-posed.  If $d_{2,2}=0,~d_{1,2}\neq 0$, we
see from \eqref{eq:fixw2} that one root will have $Re(\omega) \sim q^{2/3}$ and the gauge is again ill-posed. If
$d_{2,2}=0,~d_{1,2}= 0,~d_{2,1}\neq 0$, then one root has $Re(\omega)\sim q^{1/2}$ and the gauge is  ill-posed as well. 
Having 
$d_{2,2}=d_{1,2}=d_{2,1}= 0$ and, thus, the asymptotic behavior $\omega\sim O(1)$ for all four roots is impossible. A reasoning
similar to that used for fixed algebraic gauges above shows that this  requires $\beta_i$ to be functions of the
metric and contradicts to the assumption
\eqref{eq:metdep}.  We finally conclude that  gauges \eqref{eq:metdep} are ill-posed unless $D^{ij}e_ie_j$ is strictly
positive, i.e., unless
\begin{equation}\llabel{eq:metdep-posed}
D^{ij}e_i e_j  = \left( \beta^i\beta^j + \pd{\alpha^2}{\gamma_{ij}} \right)  e_i e_j > 0 
\end{equation}
is  satisfied for every $e_i$.

One simple gauge of type \eqref{eq:metdep} which has been used in a variety of works is a gauge which depends on the
determinant  of a three-metric, 
\begin{equation}\llabel{eq:trace}
\alpha = \alpha(x^a,\gamma) ~, ~~~\gamma =det(\gamma_{ij})~.  
\end{equation}
 For this  gauge
\begin{equation}\llabel{eq:Dijtrace}
D^{ij} = \beta^i\beta^j + \pd{\alpha^2}{\gamma} \gamma \gamma^{ij}~
\end{equation}
 ($d\gamma = \gamma \gamma^{ij} d\gamma_{ij}$), and, since $\gamma^{ij}$ is positive definite, 
the gauge will be ill-posed unless 
\begin{equation}\llabel{eq:trace-posed}
\pd{\alpha^2}{\gamma} > 0  ~.
\end{equation}

More general metric-dependent gauges may be investigated in a similar way. For example, gauges with both $\alpha$ and $\beta_i$
functions of the metric,
\begin{equation}\llabel{eq:ab}
\alpha=\alpha(x^a,\gamma_{ij})~,~~~\beta_i=\beta_i(x^a,\gamma_{ij})~,
\end{equation}
have both $U\neq 0$ and $V_i\neq 0$,  
\begin{equation}
\begin{split}
U  &= 2 \pd{\alpha}{\gamma_{ij}} \left( \Gamma^c_{ij} \psi_c  - \pd{\psi_i}{x^j}   \right)~,\\
V_i  &= 2 \pd{\beta_i}{\gamma_{jk}} \left(  \Gamma^c_{jk} \psi_c -  \pd{\psi_j}{x^k}   \right)~,\\
\end{split}
\end{equation}
and the resulting coordinate perturbation equations become more complicated
\begin{equation}\llabel{eq:abgauge}
\begin{split}
&\pd{\psi_0}{t} =  - D^{ij} \pd{\psi_i}{x^j} + C^a  \psi_a   ~, \cr
&\pd{\psi_i}{t} =  - \pd{\psi_0}{x^i} - G^{jk}_i \pd{\psi_j}{x^k} + E^a_i \psi_a~.     \cr
\end{split}                                                                                 
\end{equation}
Here
\begin{equation}\llabel{eq:coeffab}
\begin{split}
&C^a = \Gamma^a_{00} + D^{ij} \Gamma^a_{ij}~,  \\
&D^{ij} = \beta^i\beta^j +  \pd{\alpha^2}{\gamma_{ij}} - 2 \beta^n \pd{\beta_n}{\gamma_{ij}}~, \\
&E^a_k = 2 \Gamma^a_{0k} - 2 \Gamma^a_{ij}  \pd{\beta_k}{\gamma_{ij}}~, \\
&G^{ij}_k = -2  \pd{\beta_k}{\gamma_{ij}}~.\\
\end{split}
\end{equation}
The matrix $\hat M_0$  remains the same, but $\hat M_1$ now becomes
\begin{equation}
\hat M_1 = \left( \matrix
                    0    &D^{1k}e_k & D^{2k}e_k & D^{3k}e_k\\
                     e_1 &G^{1k}_1e_k &G^{2k}_1e_k &G^{3k}_1e_k\\
                     e_2 &G^{1k}_2e_k &G^{2k}_2e_k &G^{3k}_2e_k\\
                     e_3 &G^{1k}_3e_k &G^{2k}_3e_k &G^{3k}_3e_k\\
                  \endmatrix
           \right)~.
\end{equation}
The dispersion relation is still a quartic equation \eqref{eq:fixdisp1}  but
 its coefficients now are
forth-order polynomials in $q$. The  analysis of  the roots $\omega_{1,2,3,4}$ is 
a  more complicated algebraic problem. In general, \eqref{eq:ab} may be either ill- or well-posed depending on the
functional form of $\alpha$ and $\beta_i$.

It must be stressed here that a  well-posedness is not a guarantee of gauge stability.
Metric-dependent gauges satisfying \eqref{eq:metdep-posed} are generally unstable with the increment of instability
$Re(\omega) =
\omega_0$  given by formulas \eqref{eq:fixw0} with coefficients determined according to \eqref{eq:Cafix}, \eqref{eq:fixEak},
\eqref{eq:fixdij},   and
\eqref{eq:Dijmetdep}. In particular, it can be shown using these formulas that the gauge \eqref{eq:trace} can be unstable for a
wide range of background solutions.  
   When  \eqref{eq:ab} is  well-posed, it can be both stable and unstable depending on a particular background solution.

A class of gauges often considered in the literature is \cite{BONA-97}
\begin{equation}\llabel{eq:BMgauge}
\pd{\alpha}{t} - \beta^i\pd{\alpha}{x^i} = -\alpha^2 f(\alpha) \,tr( K_{ij})
\end{equation}
where $f(\alpha)$ is an arbitrary function. It was found that $f \geq 0$ is a
necessary condition for hyperbolicity of first-order 3+1 formulations of GR introduced in that paper.
 The gauge  \eqref{eq:BMgauge} is, in fact, equivalent to  an algebraic gauge \cite{ARBONA-99}
\begin{equation}
\sqrt{\gamma} = F(\alpha)
\end{equation}
with $f = \alpha F \left( \pd{F}{\alpha}\right)^{-1}$, and contains as its members 
such gauges as a harmonic slicing ($f=1$) and a "1+log" slicing ($f=1/\alpha$). 
One can see that the condition  $f > 0$ derived from the analysis of hyperbolicity of the entire 3+1 system of
equations is equivalent to  our condition of well-posedness 
$\pd{F}{\alpha} >0$ derived from the analysis of gauge modes alone.  
 
 A more general family of first-order 3+1 systems has been derived in \cite{KIDDER-01} using  a  gauge
\begin{equation}\llabel{eq:densitized}
\log(\alpha g^{-\sigma}) = Q(x^a)~,~\beta_i=\beta_i(x^a)~.
\end{equation} 
This gauge belongs to a family of algebraic gauges \eqref{eq:metdep} as well. It was found in \cite{KIDDER-01} that having a
metric-dependent, densitized  lapse with
$\sigma > 0$ is a necessary condition for a hyperbolicity of 3+1 systems considered in that work. Again, it is easy to see
that the requirement $\sigma > 0$ is equivalent to the condition of well-posedness 
 derived in this paper from the analysis of gauge instabilities.  


\subsection{Differential gauges}

As an example of a differential gauge, we consider a parabolic extension of a well known maximal slicing 
gauge $\gamma^{ij} \nabla_i \nabla_j
\alpha=K^{ij}K_{ij}\alpha~, ~\beta_i=0$ \cite{SMARR-77-1,SMARR-77-2},
\begin{equation}\llabel{eq:MaxDif}
 \pd{\alpha}{t} = {1\over \epsilon} \left(  \gamma^{ij} \nabla_i \nabla_j \alpha - K^{ij}K_{ij}\alpha \right),~\beta_i=0~,
\end{equation}
where $\epsilon >0$ is a constant. We follow the same general procedure as that used above for fixed
and algebraic gauges. Expanding  covariant derivatives in \eqref{eq:MaxDif} and taking into account that, for $\beta_i=0$, the
extrinsic curvature
$K_{ij} = - {1\over 2\alpha}
\pi_{ij}$, we rewrite \eqref{eq:MaxDif} as  
\begin{equation}\llabel{eq:MaxDif2}
F_0 \equiv -\epsilon \pd{\alpha}{t} + \gamma^{ij}  \pddd{\alpha}{x^i}{x^j} - \gamma^{ij} \lambda^k_{ij}\pd{\alpha}{x^k}
  - { \pi^{ij} \pi_{ij} \over 4\alpha}  =0~,~\beta_i=0~,
\end{equation}
and find the  derivatives of $F_0$, 
\begin{equation}\llabel{eq:MaxDif1}
\begin{split}
& \pd{F_0}{\alpha} = {\pi_{kl} \pi^{kl} \over 4\alpha^2}~, ~\pd{F_0}{(\pd{\alpha}{x^i})} = -\gamma^{kl} \lambda^i_{kl}~,~
\pd{F_0}{(\pddd{\alpha}{x^i}{x^j})} = \gamma^{ij}~,~\pd{F_0}{(\pd{\alpha}{t})} = -\epsilon~,\\ 
& \pd{F_0}{\gamma_{ij}} \equiv A^{ij} = {\gamma_{kl}\, \pi^{ik}\pi^{jl}\over 2\alpha}  
        - \gamma^{ik} \gamma^{jl} \pddd{\alpha}{x^k}{x^l} + \left( \gamma^{ik} \gamma^{jl} \lambda^n_{kl}
        +  \gamma^{kl} \gamma^{in} \lambda^j_{kl} \right) \pd{\alpha}{x^n} ~,\\
& \pd{F_0}{(\pd{\gamma_{ij}}{x^k})} \equiv B^{ij,k} = - {1\over 2} \pd{\alpha}{x^n} \left( \gamma^{ik} \gamma^{jn} +
             \gamma^{jk}\gamma^{in}  - \gamma^{ij}\gamma^{kn} \right)~,~ \text{and}~
 \pd{F_0}{\pi_{ij}}   = - {\pi^{ij}\over 2\alpha} ~.    \\
\end{split}
\end{equation}
Substituting \eqref{eq:MaxDif1}  into \eqref{eq:CaseDifferential}, and combining \eqref{eq:CaseDifferential} with
\eqref{eq:gauge} we obtain the system of five quasi-linear partial differential equations
\begin{equation}\llabel{eq:MaxDif3}
\begin{split}
\pd{\psi_0}{t} =& \, \alpha U  + \Gamma^c_{00}  \psi_c   ~, \\
\pd{\psi_i}{t} =&  - \pd{\psi_0}{x^i} + 2\Gamma^c_{0i} \psi_c~,  \\
\epsilon \pd{U}{t}      =& \, \gamma^{ij} \nabla_i \nabla_j U + {\pi_{ij} \pi^{ij}\over 4\alpha^2}  U   -
{\pi^{ij}\over\alpha} 
\pddd{\psi_0}{x^i}{x^j} - 2 B^{ij,k} \, \pddd{\psi_i}{x^j}{x^k} 
             +   C^i \pd{\psi_0}{x_i} + D^{ij}  \pd{\psi_i}{x_j} + J^c\psi_c ~,\\
\end{split}   
\end{equation}
where
\begin{equation}\llabel{eq:Maxcoef}
\begin{split}
&C^k =  {\pi^{ij}\over\alpha} \Gamma^k_{ij} + 2 {\pi^{ik}  \over\alpha} \Gamma^0_{0i} + 2 B^{ij,k} \Gamma^0_{ij} ~,~~D^{ij} = 
{2 \Gamma^i_{0k} \pi^{kj}\over\alpha} - 2 A^{ij} + 2 B^{lk,j} \Gamma^i_{lk} ~, ~\text{and}\\ &J^c = 2 A^{ij} \Gamma^c_{ij}  -
{\pi^{ij}\over\alpha}   \left( \pd{\Gamma^c_{ij}}{t} + \Gamma^0_{ij} \Gamma^c_{00} + 2
\Gamma^k_{ij} \Gamma^c_{0k} -2\pd{\Gamma^c_{0i}}{x^j} \right)  + 2 B^{ij,k}  \pd{\Gamma^c_{ij}}{x^k}  ~. \\
\end{split}
\end{equation}
The dispersion relation becomes
\begin{equation}\llabel{eq:MaxMatrix}
det\, \left( \matrix 
                            & \Gamma^0_{00} - \omega   &\Gamma^1_{00} &\Gamma^2_{00} &\Gamma^3_{00}  &\alpha & \\
                                & & & &\\
                            & 2\Gamma^0_{01}+\ImI e_1 q    &2\Gamma^1_{01}-\omega &2\Gamma^2_{01} &2\Gamma^3_{01}  &0 &\\
                                & & & &\\
                            & 2\Gamma^0_{02}+\ImI e_2 q    &2\Gamma^1_{02} &2\Gamma^2_{02} -\omega &2\Gamma^3_{02}  &0 &\\
                                & & & &\\
                            & 2\Gamma^0_{03}+\ImI e_3 q    &2\Gamma^1_{03} &2\Gamma^2_{03} &2\Gamma^3_{03} -\omega  &0 &\\
                                & & & &\\
                            & M_{40}    & M_{41} & M_{42} & M_{43}  &  M_{44} - \omega &\\
                        \endmatrix \right) =0 ~,
\end{equation}
where
\begin{equation}\llabel{eq:MaxMCoeff}
\begin{split}
&M_{40} = {1\over\epsilon}\left(J^0-\ImI C^ie_i q + {\pi^{ij}e_ie_j\over\alpha} q^2 \right) ~,\\
&M_{4i} = {1\over\epsilon}\left( J^i-\ImI D^{ij}e_j q + 2 B^{ij,l} e_je_l q^2 \right) ~,~\text{and} \\
&M_{44} = {1\over\epsilon} \left( {\pi_{ij} \pi^{ij}\over 4\alpha^2} + \ImI \gamma^{ij}\lambda^k_{ij} e_k\, q
- \gamma^{ij}e_ie_j\,q^2  \right)~.\\
\end{split}
\end{equation}
The dispersion relation \eqref{eq:MaxMatrix} is a fifth-order equation with respect to $\omega$,
\begin{equation}\llabel{eq:MaxDisp}
-\omega^5 + d_4 \omega^4 + d_3\omega^3 + d_2\omega^2 + d_1\omega + d_0 = 0~,
\end{equation}
where
\begin{equation}\llabel{eq:MaxDispCoeff}
\begin{split}
& d_4 = d_{4,0} + d_{4,1} q + d_{4,2} q^2~,\\
& d_3 = d_{3,0} + d_{3,1} q + d_{3,2} q^2~,\\
& d_2 = d_{2,0} + d_{2,1} q + d_{2,2} q^2 + d_{2,3} q^3~,\\
& d_1 = d_{1,0} + d_{1,1} q + d_{1,2} q^2 + d_{1,3} q^3~,\\
& d_0 = d_{0,0} + d_{0,1} q + d_{0,2} q^2 + d_{0,3} q^3~,\\
\end{split}
\end{equation}
and coefficients $d_{i,j}$ are functions of an unperturbed metric $g_{ab}$.
In what follows, we need explicit expressions for 
\begin{equation}\llabel{eq:MaxDispCoeff1}
\begin{split}
 d_{4,2} = &\, \epsilon^{-1} \gamma^{ij}e_ie_j  ,\\
 d_{4,1} = &\,  \ImI \epsilon^{-1} \gamma^{ij}\lambda^k_{ij} e_k~,\\
 d_{2,3} = &\, \ImI \left( 2 \epsilon^{-1}  \alpha B^{ij,k} e_i e_j e_k +  d_{4,2} \Gamma^k_{00} e_k \right) = - \ImI
d_{4,2} \Gamma^k_{00} e_k = -\ImI \alpha \pd{\alpha}{x^n} d_{4,2} \gamma^{nk} e_k~.\\
\end{split}
\end{equation}
From \eqref{eq:MaxDisp}, \eqref{eq:MaxDispCoeff} we obtain the following asymptotic behavior of
the roots.
\begin{equation}\llabel{eq:Maxroot}
\begin{split}
\omega_1 &= - d_{4,2} \, q^2 - d_{4,1} q + O\left(1\right)~, \\
\omega_{2,3} & =  \pm  q^{1/2} \sqrt{ - {d_{2,3} \over  d_{4,2}}  } + O\left( 1\right) 
               = \pm \left(1- \ImI \,{\Gamma^k_{00} e_k \over \vert \Gamma^k_{00} e_k \vert} \right) q^{1/2} \sqrt{ \vert \Gamma^k_{00}
e_k
\vert
\over 2}+ O\left( 1\right) ~,\\
\omega_{4,5} & =  O(1)~. \\
\end{split}
\end{equation}
Since $\gamma^{ij}$ is positive definite, we have $d_{4,2} >0$, and thus $Re(\omega_1) < 0$ in the limit of $q\rightarrow \infty$.
This root does not cause an instability. However, one of the roots $\omega_{2,3}$  has a positive $Re(\omega) $ growing with $q$
as
$$
Re(\omega) \propto q^{1/2}\, \sqrt{\Gamma^k_{00} e_k \over 2}  = q^{1/2} \,\sqrt{\frac{1}{2} \alpha \gamma^{ik} e_k \bigg|
\pd{\alpha}{x^k} \bigg| }
 ~.
$$
Since the direction of $e_i$ is arbitrary,  a maximal slicing gauge will be ill posed unless $\alpha$ is spatially constant. The
result we obtain for a maximal slicing and its parabolic extension is exactly the same  as that we found earlier for
fixed algebraic gauges  (see
\eqref{eq:fixzero}). It is easy to see that in both cases the asymptotic $q^{1/2}$-behavior of  the
roots  comes from the presence of 
$\Gamma^i_{00} \psi_i$ term in the right-hand side of the first equation in either \eqref{eq:fixedgauge} or \eqref{eq:MaxDif3}.
Differentiating  first equation with respect to time and substituting  second equation into it, we obtain in both cases a parabolic-like
equation for $\psi_0$
$$
\pdd{\psi_0}{t} = \Gamma^i_{00}\pd{\psi_0}{x^i} + {\rm other~terms}
$$
which is ill-posed if initial conditions are specified at $t=const.$ (see also next section).


\section{Physical meaning of gauge instabilities}

\subsection{Gauge transformations in a flat spacetime}

To illustrate our analysis of gauge stability, we must consider now the entire evolution part of the Einstein equations and
compare its stability with that of the system \eqref{eq:gauge}, \eqref{eq:CaseDifferential}.  We will use an  ADM 3+1 system with 
metric
$\gamma_{ij}$ and extrinsic curvature 
$K_{ij}$ as unknown variables,
\begin{equation}\llabel{eq:ADM}
\begin{split}
{\partial\gamma_{ij}\over\partial t} &= -2\alpha K_{ij} + \nabla_j\beta_i + \nabla_i\beta_j ~,\\
{\partial K_{ij}\over\partial t} &= -\nabla_i\nabla_j\alpha +\alpha\left( R_{ij} + K K_{ij} - 2
K_{im} K^m_j \right)\\ 
& +\beta^m\nabla_mK_{ij} +K_{im}\nabla_j\beta^m +K_{jm}\nabla_i\beta^m~, \\
\end{split}
\end{equation}
and will consider a simple class of one-dimensional solutions of  \eqref{eq:ADM}
\begin{equation}\llabel{eq:MetricCoord1D}
\gamma_{ij} = diag(\gamma,1,1)~, K_{ij} = diag(K,0,0)
\end{equation}
with $\gamma$ and $K$ dependent on $t$ and $x\equiv x^1$ only.  
 In vector notations, \eqref{eq:ADM} becomes 
\begin{equation}\llabel{eq:ADM1D}
\pd{{\bf u}}{t} = {\bf A }({\bf u}) + \hat B ({\bf u}) \pd{{\bf u}}{x}
\end{equation}
where
\begin{equation}\llabel{eq:ADM1Da}
{\bf u} = \pmatrix \gamma\\ K\\\endpmatrix~,~~~{ \bf A }= \pmatrix 
                  -2\alpha K +2\pd{\beta}{x}\\
                  -\pdd{\alpha}{x}  - {\alpha K^2\over\gamma} +{2K\over\gamma}\pd{\beta}{x}\\
         \endpmatrix~,~~~
\hat B = \pmatrix 
 - {\beta\over\gamma} & & 0 \\             
                          {1\over 2\gamma}\pd{\alpha}{x}-{2K\beta\over\gamma^2} &
                                                               & {\beta\over\gamma} \\
         \endpmatrix~.
\end{equation}
 One can easily verify that all components of the Riemann
tensor are identically zero for any metric \eqref{eq:MetricCoord1D} satisfying \eqref{eq:ADM1D}. That is, 
\eqref{eq:MetricCoord1D} describes  a flat
spacetime in non-Galilean coordinates\footnote{The only nontrivial component of the Riemann tensor in this case is $R_{1010}$. Since
choice of
$\alpha$,
$\beta$ is a choice of a coordinate system, it is sufficient to show that $R_{1010} \equiv 0$ in a coordinate system with
$\alpha=1$, $\beta=0$. In this case, non-zero
 Cristoffels  are 
$$
\Gamma_{01}^1 = \Gamma^1_{01} = {1\over 2\gamma}{\partial \gamma\over\partial t}~,~~ 
\Gamma_{11}^0 = {1\over 2} {\partial \gamma\over\partial t}~,~~
\Gamma_{11}^1  ={1\over 2 \gamma} {\partial \gamma\over\partial x}~,
$$
and
$$
R_{0101} = -{1\over 2} {\partial^2 g_{11}\over\partial t^2}
           +g_{11} \Gamma^1_{10}  \Gamma^1_{01} 
         = -{1\over 2} {\partial^2 \gamma\over\partial t^2}
          + {1\over 4\gamma} 
              \left(\partial\gamma\over\partial t\right)^2 \equiv 0~. 
$$
}. One can also verify that the constraint equations are automatically satisfied for any
metric
\eqref{eq:MetricCoord1D}. Therefore,   \eqref{eq:ADM1D} can be
unstable only with respect to  gauge instabilities.

Eigenvectors and eigenvalues of $\hat B$ in \eqref{eq:ADM1D}  are
\begin{equation}\llabel{eq:ADM1Db}
\lambda = \pm {\beta\over\gamma}~, 
   ~{\bf u} = \pmatrix 4\beta\gamma / ( 4\beta K - \gamma \pd{\alpha}{x} ) \\ 1 \\ \endpmatrix~,
            ~~\pmatrix 0\\ 1\\ \endpmatrix~.
\end{equation}
For $\beta\neq 0$, there is a complete set of of eigenvalues and eigenvectors, so that  \eqref{eq:ADM1D} is strongly hyperbolic
and well posed. For $\beta=0$, a complete set does not exists, and the  system reduces to a parabolic equation
\begin{equation}\llabel{eq:ADM1Dc}
\pdd{\gamma}{t} = \left({\alpha\over\gamma}\pd{\alpha}{x} \right) \pd{\gamma}{x} 
                + 2\alpha^2 \, { K^2\over\gamma} - 2\pd{\alpha}{t}\, K + 2\alpha\pdd{\alpha}{x} ~
\end{equation}
which is ill-posed  (a parabolic equation $\pd{a}{\tilde x} = \pdd{a}{\tilde y}$ is ill posed
if initial conditions are defined at $\tilde x=const$, e.g.,  \cite{WILLIAMS-80}).

Analysis of the principal part $\hat B \pd{{\bf u}}{x}
$ is not sufficient, however, for investigation of instabilities of \eqref{eq:ADM1D}.
We need to  carry out a full stability analysis of this system. A linearized version of  \eqref{eq:ADM1D} is 
\begin{equation}\llabel{eq:ADM1Dd}
\pd{{\bf \bar u}}{t} =  \hat C({\bf u}) \, {\bf \bar u} + \hat B ({\bf u}) \, \pd{{\bf \bar u}}{x}  
\end{equation}
where $\bf u$ is now an unperturbed base solution, $\bf \bar u$ is a vector of perturbations, 
and
\begin{equation}\llabel{eq:ADM1De}
\hat C = \pd{{\bf A}}{{\bf u}} + \pd{\hat B}{{\bf u}}\pd{{\bf u}}{x} = \left(\matrix
                           {\beta\over\gamma^2}\pd{\gamma}{x} & & -2\alpha \\ 
                           {\alpha K^2\over\gamma^2}  
          -{\partial\alpha\over\partial x}{1\over 2\gamma^2}\pd{\gamma}{x}-
{2\beta\over\gamma^2}\pd{K}{x}
          - {2K\over\gamma^2}\pd{\beta}{x}+{4K\beta\over \gamma^3}\pd{\gamma}{x}
& &  {2\over\gamma}\pd{\beta}{x} -{\beta\over\gamma^2}\pd{\gamma}{x} 
   -{2 \alpha K\over\gamma} \\
               \endmatrix                
 \right) ~.  
\end{equation}
For harmonic solutions
\begin{equation}
{\bf \bar u}  \propto \exp ( \omega t -
\ImI qx)~,                                                                                  
\end{equation}
 in the limit  $q \rightarrow\infty$ we then obtain a  dispersion relation 
\begin{equation}\llabel{eq:ADM1Df}
\vert\vert \hat C -\ImI q \hat B - \hat I \omega \vert\vert = \omega^2 + d_1\omega + d_0 = 0~,  
\end{equation}
with
\begin{equation}\llabel{eq:ADM1Dg}
\begin{split}
d_1 &= 2 \left( { \alpha K\over\gamma} - {1\over\gamma}\pd{\beta}{x} \right)~,\\
d_0 & = d_{0,0} + \ImI d_{0,1}q + d_{0,2} q^2 ~, \\
\end{split}
\end{equation}
and 
\begin{equation}\llabel{eq:ADM1Dh}
\begin{split}
    d_{0,0} &= \left({\beta\over\gamma^2}\pd{\gamma}{x} \right) \left( {2\over\gamma}\pd{\beta}{x}
                   -{2\beta\over\gamma^2}\pd{\gamma}{x} 
                          -{2 \alpha K\over\gamma}\right)  \\
        &+ 2\alpha\left( {\alpha K^2\over\gamma^2}  
          -{\partial\alpha\over\partial x}{1\over 2\gamma^2}{\partial\gamma\over\partial x}-
{\beta\over\gamma^2}\pd{K}{x}
          - {2K\over\gamma^2}\pd{\beta}{x}+{4K\beta\over \gamma^3}\pd{\gamma}{x} \right)~, \\ 
d_{0,1} &=   {2\beta\over\gamma} \left( {1\over\gamma}\pd{\beta}{x} -{\beta\over\gamma^2}\pd{\gamma}{x} 
   +{ \alpha K\over\gamma}  \right) -{\alpha \over
\gamma}\pd{\alpha}{x} 
 ~, \\
d_{0,2} &= {\beta^2\over\gamma^2}~. 
\end{split}
\end{equation}
For 
$\beta\neq 0$, $d_{0,2} \neq 0$ and we get an asymptotic behavior
\begin{equation}\llabel{eq:ADM1Di} 
w = \pm  \ImI q\,{\beta\over\gamma}  \pm {\alpha\over 2 \beta} \pd{\alpha}{x} \pm {\beta\over \gamma^2}\pd{\gamma}{x}
- { \alpha K\over\gamma} + {1\over\gamma}\pd{\beta}{x}
    \mp \left ( { \alpha K\over\gamma} + {1\over\gamma}\pd{\beta}{x}
                           \right)  + O \left(\frac{1}{q} ~.
\right)
\end{equation}
We see that although the system is well-posed, it has $Re(\omega) \sim O(1)$, and is  unstable with respect to small
perturbations (unless
$\pd{\beta}{x}=\pd{\gamma}{x}=\pd{\alpha}{x}=0$, and $K>0$). If
$\pd{\alpha}{x}
\neq 0$,  the increment of the instability $Re(\omega)\sim \beta^{-1} \rightarrow\infty$ when $\beta \rightarrow 0$.  

For 
$\beta=0$ and $\pd{\alpha}{x}\neq 0$, we have $d_{0,2} =0$, $d_{0,1}\neq 0$,  and the asymptotic behavior is
\begin{equation}\llabel{eq:ADM1Dj}
\omega = \pm (1 + \ImI ) \,q^{1/2}  \sqrt{{\alpha\over 2\gamma} \pd{\alpha}{x}} + O(1)
\end{equation}
which 
shows that  $Re(\omega) \sim q^{1/2}$ and the system is  ill-posed.  The conclusions agree with the results of gauge stability
analysis of the previous section.


\subsection{Instability of a synchronous coordinate system}

   Synchronous
coordinate systems are formed by acceleration-free test particles. It is well known that in such systems the metric
determinant $\vert\gamma_{ij}\vert$
 will vanish after a finite time because the time-lines of a reference frame will necessarily
intersect one another on a certain caustic hypersurface \cite{LL}. This is correct in both flat and curved spacetimes.
 On a
caustic,  one of the principal values of the metric tensor vanishes,  whereas the corresponding contravariant component tends to
infinity. In the vicinity of a caustic, using an arbitrariness in the selection of spatial coordinates
one can write a general four-metric  as \cite{LIFSHITS-61}
\begin{equation}\llabel{eq:CAUSTIC}
g_{00} = -1~,~~g_{0i} = 0~, ~~g_{np} = \gamma_{np} = a_{np}~, ~~g_{n3} = \gamma_{n1} = \tau^2 a_{n1}~,~~g_{11}=\gamma_{11} =
\tau^2 a_{11}
\end{equation}
where $\tau= t-x^1$, indices $n,p$ take values $2,3$, and $a_{ij}$ are non-singular functions of $x^2$, $x^3$, and $t$.
Coefficients $a_{ik}$ are connected by a single relation which is a consequence of the Einstein equations (see Appendix in
\cite{LIFSHITS-61}). Equation \eqref{eq:CAUSTIC} determines a general ``quadratic" character of approaching a caustic in a 
synchronous gauge.
In \eqref{eq:CAUSTIC}, some of the contravariant components of $g^{ab}$ tend to infinity as $\tau^{-2}$. Because of that, the process
described by \eqref{eq:CAUSTIC} is sometimes called  a ``blow up" instability.  

It is important to distinguish between the formation of a caustic (blow up) and the   instability of a synchronous gauge
 with respect to small perturbations considered in this paper. To clarify this important point, let us consider the following example.
Consider  one-dimensional  solutions of \eqref{eq:ADM1D} for $\alpha=1$ and $\beta=0$. The  equations
\eqref{eq:ADM1D} become
\begin{equation}\llabel{eq:Sync1D}
\pd{\gamma}{t} = -2K~, ~~ \pd{K}{t} = -{K^2\over\gamma}~,
\end{equation}
and a general solution of \eqref{eq:Sync1D} is\footnote{Physical meaning of $A$ and $B$ in this solution is discussed in the Appendix} 
\begin{equation}\llabel{eq:exactsync}
\gamma = (A(x) + B(x) t )^2~, ~~~K =  - B(x) ( A(x) + B(x) t ).
\end{equation} 
If at some point we have $A>0$, $B<0$,  
a caustic will form and both $\gamma$ and $K$ will become zero  at some future time $t_c= - A/B >0$.
Note that the process of
caustic formation may be global if $\pd{A}{x},\pd{B}{x} = 0$ or local otherwise. 
 Using
\eqref{eq:exactsync} it is easy to write the deviation of $\gamma$, $\delta\gamma$ as a function of perturbations of
$A$ and
$B$,
\begin{equation}
\delta\gamma = 2 (A + B t ) ( A \,\delta A + B \,\delta B \,t )~,
\end{equation}
where $\delta A = \Delta A / A$, $\delta B = \Delta B / B$, and $\delta A, \delta B << 1$. This is a weak instability  described by a
power function of $t$ instead of an exponential function\footnote{A formal stability analysis gives in this case a growth rate of
perturbations $\omega = - K/\gamma$ comparable to that of an unperturbed solution. As it was discussed in  Section 3,
\eqref{eq:Harmpert} is not  valid in this case.}. 
  Both the formation
of a caustic and the instability are due to the presence of  initial velocities of the reference frame with respect to an inertial
Galilean frame. However, the caustic is  due to the intersection of  converging trajectories of  test particles, whereas the instability
refers to 
 a divergence of  test particles from unperturbed time-lines.

\subsection{Gauge instabilities in accelerating systems}

Let us now discuss a physical meaning of ill-posedness and instability of  non-synchronous gauges.
Consider first a Rindler reference frame \cite{RINDLER-66} where this meaning is most clear. 
Let $T,X^1,X^2,X^3$ be  a Cartesian coordinate system in a Minkowski space-time with an interval 
\begin{equation}
ds^2 = - dT^2 + (dX^1)^2 + (dX^2)^2 + (dX^3)^2
\end{equation}
 The Rindler
frame is constructed from the world lines of particles moving with accelerations along an X-axis of this coordinate system
in such a way that the accelerations are 
 constant in proper time of the frame. Such a frame  is "rigid" in the sense that It does not deform in proper
time. 
For an accelerating  frame to be rigid, accelerations of the particles  must be different with respect to
the inertial frame. If accelerations of particles with different $X^1$ and same time $T$ are equal, the distance between
particles measured in the inertial coordinate frame will stay constant.  Due to a Lorentz contraction, distances
measured in the accelerating frame will depend on the velocity of the frame, and will change with time.

The Rindler frame can be described  using  coordinates $x,t$ (we omit two other spatial coordinates for brevity) as
\begin{equation}\llabel{eq:Rindler}
T = x \sinh (gt),~ X^1 = x \cosh (gt)~,
\end{equation}
where  $g$ is a constant.  In these coordinates
the interval  becomes
\begin{equation}
ds^2 = -(gx)^2 dt^2 + dx^2 ~,
\end{equation}
and it is evident that the geometry of the co-moving space is this frame is time-independent.   Coordinate lines of $t$, $
T^2 = X^2 - x^2$,
 are world lines of uniformly accelerated particles. An  acceleration $a$ measured
by a co-moving observer in an orthogonal coordinate system $x^a$ is \cite{FN}
\begin{equation}
a = \sqrt{\gamma_{ik} a^i a^k }~,~~a^i = \Gamma^i_{00} / g_{00} ~.
\end{equation}
For a system with acceleration along the $X^1$-axis we obtain 
\begin{equation}\llabel{eq:Accel}
a = {1\over\alpha \sqrt{\gamma_{11}}} \left| \pd{\alpha}{x}\right|
\end{equation}
and for a Rindler frame
\begin{equation}\llabel{eq:Rindler2}
a_{R} = {1\over x} 
\end{equation}
In a Rindler frame, the coordinate $x$ is equal to a physical distance $l=\int_0^x \sqrt{\gamma_{11} } dx$ from the point
$x=0$ where physical acceleration is infinite. Therefore, \eqref{eq:Rindler2} can be written in terms of quantities which
have a direct physical meaning,
\begin{equation}\llabel{eq:Rindler3}
a_R = {1 \over l}~.
\end{equation}
The dependence \eqref{eq:Rindler3} of a physical acceleration on a physical distance  provides the rigidity of a Rindler frame.

The perturbation equations \eqref{eq:ADM1Dd} for a Rindler frame can be reduced to a single equation
\begin{equation}\llabel{eq:Rindler4}
\pdd{\bar\gamma}{t} = -g^2 x \pd{\bar\gamma}{x}
\end{equation}
where $\bar\gamma$ is a perturbation of $\gamma_{11}$. 
Consider the following exact solutions of \eqref{eq:Rindler4} (modes of perturbation):

\begin{enumerate}
\item $\bar \gamma = const << 1$. This corresponds to a uniform change of the length scale along the $x$-axis. Such a
perturbation changes both the acceleration $a={1\over x} (1 + \bar\gamma)^{-1/2}$ and the physical distance $l=\int_0^x
(1+\bar\gamma)^{1/2}dx =  (1+\bar\gamma)^{1/2} x$ in such a way that still $a=1/l$. The condition \eqref{eq:Rindler2}  is not
violated and the  frame remains rigid. 
\item $\bar \gamma = A t$, $At << 1$. The perturbation does not depend on $x$ and describes a deformation of the
frame with constant velocities with respect to a Rindler frame.  Formally this is an instability but
 of the type similar to that existing in synchronous gauges. As in the previous case, it can be shown that 
 the relation $a=1/l$ holds.
\item $\bar\gamma = x^{-A^2/g^2} \left( c_1 \sinh(At) + c_2\cosh(At) \right)$. This time the initial perturbation depends on $x$
and grows exponentially. Suppose $c_2>0$ so that $\bar\gamma >0$, then
$\pd{\bar\gamma}{x} < 0$.  The acceleration is again $a(x) = {1\over x} (1+\bar\gamma(x))^{-1/2}$. The physical distance,
however, is 
 $l=\int^x_0(1+\bar\gamma)^{1/2}dx = x ( 1 + \bar\gamma( x' ))^{1/2} $ with some $ x'<
x$. As a result, 
$a(x) > 1/l(x)$ is now greater than that required to maintain rigidity, and the deformation will grow with time.
 A perturbation $\bar\gamma = x^{A^2/g^2} \left( c_1 \sin(At) + c_2\cos(At) \right)$ is also a solution of \eqref{eq:Rindler4},
but now $\pd{\bar\gamma}{x} < 0$ and this particular perturbation is stable.
\end{enumerate}
   
We see that the reason for the instability is a mismatch between the deformation of the frame and the
prescribed acceleration of particles which leads to further frame deformation. We can imagine, for example, a deformation
with $\bar\gamma=0$ everywhere for $x$ less than some value $x_0$, $\bar\gamma = \bar\gamma_0 =const > 0$ at $x>x_1>x_0$ and
$\bar\gamma$ increasing linearly between $x_0$ and $x_1$.  The difference between
accelerations at $x_0$ and $x_1$ will be finite, but by tending $x_1 \rightarrow x_0$ we can make $\gamma(t)/\gamma(0)$ between
the neighboring points $x_0$ and $x_1$ to increase with any desired rate. This corresponds
to the mathematical property of ill-posedness of fixed gauges with spatially non-uniform lapse. It
is also obvious that when $\alpha$ and $\gamma$ depend on $t$ and $x$, the instability will mean a deviation from  an already
deforming  reference frame. 

In one-dimensional case, any metric with non-zero shift can be obtained from a metric with $\beta=0$ by a transformation
of a time coordinate $\bar t = \bar t(t,x)$. Such a transformation does not change the accelerations of  test particles
which remain ``chronometric invariants" \cite{ZELMANOV-56}.  However, due to a non-zero shift,
perturbations are now ``advected", that is,  their coordinates change continuously whereas accelerations are still
prescribed at  fixed coordinates. As a result, $\gamma(x)$
cannot grow with an arbitrary rate. An apparent growth of the instability at fixed $x$ will decrease with increasing
$\beta$.  In our stability analysis, this  corresponds to a changes from an ill-posed to a well-posed but
unstable  gauge. 


\subsection{Instability of rotating reference frames}
 
In one-dimensional cases considered above the cause of instability and ill-posedness was the perturbation of
accelerations of test particles forming the reference frame due to perturbation of  positions of these
particles.
In more than one dimensions, there is another physical factor, rotation and Coriolis
forces associated with it. However, as we will see below, this factor does not change the nature of instability.   Consider, for
example, a uniformly rotating reference frame
\begin{equation}\llabel{eq:Coriolis}
\alpha = 1, ~ \beta_i = \{ -\Omega x^2, \beta_2 = \Omega x^1,0\}, ~\gamma_{ij} = diag(1,1,1)~.
\end{equation}
where $\Omega$ is an angular velocity. For \eqref{eq:Coriolis}  we have 
\begin{equation}
C^i = \Gamma^i_{00} + \beta^j\beta^k \Gamma^i_{jk} = \Gamma^i_{00} = \{\Omega^2 x^1, \Omega^2 x^2, 0 \}~,
\end{equation}
and, according to our gauge stability analysis \eqref{eq:fixCzero}, a perturbation with a wave vector $ q_i = q e_i$ 
will 
grow with the increment
\begin{equation}\llabel{eq:rotinc}
Re(w) = \Omega \,q^{1/2} \, \sqrt{ x^1 e_1 + x^2 e_2} ~~. 
\end{equation} 
Note, that 
$C^i$ are proportional to components of physical acceleration $a_i$ in \eqref{eq:Accel}.  The increment $\propto q^{1/2}$
indicates that the gauge is ill-posed.   The reason is the same as in a one-dimensional case. It is related to a radial acceleration
and not
to Coriolis force. Note that at the axis of rotation $x^1=x^2 =0$ we have $C^i=0$ and there is no instability ($Re(\omega)=0$ in
\eqref{eq:rotinc}), but of course, there is a Coriolis force. Thus we conclude that in a three-dimensional case with rotation 
the main physical reason for gauge instability and ill-posedness is the same acceleration $a^i$.


\section{Conclusions}

In this paper we presented a general approach  to the analysis of gauge stability of 3+1 formulations of GR.  Gauge modes
of perturbations can be separated, in a linear approximation, from other modes of perturbations and  studied
independently. 
 A system of
eight  quasi-linear partial differential equations \eqref{eq:gauge}, \eqref{eq:CaseDifferential}  describes the
evolution of  coordinate perturbations  $\psi_a$ and the corresponding perturbations of lapse and shift, $U$ and $V_k$, with
time. The gauge stability  depend on the choice of gauge and on  an unperturbed four-metric $g_{ab}$, 
but it does not depend on a particular form of a 3+1 system of GR equations. 

Well-posedness and stability of several gauges was investigated. We demonstrated that all fixed gauges,
i.e., gauges that are functions of coordinates only, are ill-posed with the  exception of a synchronous gauge
$\alpha=\alpha(t)$, $\beta_i=0$. This gauge is well-posed, but it is  prone to the formation of coordinate
singularities (caustics) and it is unstable. 

It is known that maximal slicing $tr(K_{ij})=0$ prevents the
formation of coordinate singularities  by applying accelerations to test particles forming a reference frame in
such a way that
$\pd{\gamma}{t}
\propto tr(K_{ij}) = 0$ and the local volume element remain
$\gamma^{1/2} = const.$ \cite{SMARR-77-2}. This, and a singularity-avoiding properties of  maximal slicing allow in many cases  a 
much longer numerical integration than using a synchronous gauge.  However,
both the maximal slicing gauge and its parabolic extension
\eqref{eq:MaxDif} are ill-posed due to the presence of acceleration-related unstable modes. Computations using these gauges will
 blow  after a long enough period of integration.  

Stability of metric-dependent algebraic gauges has been investigated as well. In particular, the necessary condition of
well-posedness of  gauges with metric-dependent lapse and fixed shift \eqref{eq:metdep} was formulated. In addition to the
formation of caustics,  algebraic well-posed gauges with spatially dependent lapse and shift are susceptible to
instabilities caused by perturbations of accelerations in deforming reference frames. The reason for the unstable behavior of
these hyperbolic gauges is the same as that for ill-posedness of fixed, parabolic, or
elliptic gauges. However, due to an ``advective" property of hyperbolic gauges, the growth rate of acceleration-related unstable modes
becomes wavelength-independent, 
$Re(\omega) \sim O(1)$, in the limit $q\rightarrow \infty$.

An investigation of stability of 3+1 formulations of GR  in this paper is limited to an investigation of gauge instabilities.
By studying
ill-posedness and stability  of
\eqref{eq:gauge},
\eqref{eq:CaseDifferential} one can tell if a 3+1 sets of GR
equations using a particular   gauge will be ill-posed or unstable. All gauges studied in this paper were found either ill-posed or
unstable. None of the gauges investigated in this paper are suitable for a long-term stable integration of GR equations. 
Gauges with better stability properties must be found. 
 
 Stability of a gauge does not mean, however, that a 3+1 system using this gauge will be  stable. As was
mentioned in the introduction,  another  source of ill-posedness and instability  is  associated with
violation of constraints. 
Constraint instabilities do depend on a particular form of a 3+1 system. The analysis of hyperbolicity in \cite{KIDDER-01} provides an
illustration of this statement. A part of the hyperbolicity conditions derived in this work (their equation (2.36)) does not involve
gauge at all, and is dependent on how the constraint equations are incorporated into  the  system. 

\bigskip\leftline{\bf Acknowledgments}
\medskip
\noindent
This work was supported in part by the NASA grant SPA-00-067, Danish Natural Science Research Council through grant No 94016535, 
Danmarks
Grundforskningsfond through its support for establishment of the Theoretical Astrophysics Center, and by the Naval Research Laboratory
through the Office of Naval Research. The authors thank A. Doroshkevich, N. Khokhlova, M. Kiel, L. Lindblom, M. Scheel, K. Thorne, and 
M. Vishik for stimulating discussions and useful comments. I.D. thanks the Naval Research Laboratory, A.K. thanks the
 Theoretical Astrophysics Center,  and both authors thank Caltech for hospitality
during their visits.
\begin{appendix}
\end{appendix}

\clearpage

\leftline{\bf Appendix: Physical meaning of $A$ and $B$ in \eqref{eq:exactsync}}

\renewcommand{\theequation}{A.\arabic{equation}} \setcounter{equation}{0}

\bigskip\noindent
Let us introduce the following new coordinates $\tilde t$, $\tilde x^1$, $\tilde x^2$, $\tilde x^3$, 
\begin{equation}
\begin{split}
& T = {\tilde t + U \tilde x^1\over \sqrt{1-U^2}}~, \\
& X^1 = {\tilde x^1 + U \tilde t \over \sqrt{1-U^2}} + \int_0^{\tilde x^1}   \left(\pd{U}{y} \right){y dy \over U \sqrt{1-U^2}}~, \\
& X^2 = \tilde x^2~,~~X^3 = \tilde x^3~, \\
\end{split}
\end{equation}
where $T,X^i$ are  Cartesian coordinates in a Minkowski spacetime $g_{ab} = diag(-1,1,1,1)$.
In these new coordinates we get the metric with the components
\begin{equation}
g_{00} = -1~,~g_{11} = \left(  A + B t  \right)^2~,~g_{22}=g_{33}=1~,~\text{others}~g_{ik}=0~, 
\end{equation}
where
\begin{equation}
A = 1 +   {\tilde x^1\over U (1-U^2)}\left( \pd{U}{\tilde x^1}\right)~,~B = 
                {1\over 
(1-U^2)}\left( \pd{U}{\tilde x^1}\right)~.
\end{equation}
It is clear that (A.1) is a generalization of a Lorentz transformation with the velocity $U$ depending on $\tilde x^1$.


\clearpage
\begin{thebibliography}{10}

\bibitem{ADM}
R.~Arnowitt, S.~Deser, and C.W. Misner.
\newblock {\em Gravitation: An Introduction to Current research}, pages
  227--265.
\newblock Wiley, New York, 1962.

\bibitem{BONA-89}
C.~Bona and J.~Masso.
\newblock Einstein's evolution equations as a system of balance laws.
\newblock {\em Phys. Rev.}, D 40:1022--1026, 1989.

\bibitem{ABRAHAMS-95}
A.~Abrahams, A.~Anderson, Y.~Choquet-Bruhat, and Jr. J.~W.~York.
\newblock Einstein and {Y}ang-{M}ills theories in hyperbolic form without gauge
  fixing.
\newblock {\em Phys. Rev. Lett.}, 75:3377­3381, 1995.

\bibitem{FRITTELLI-96}
S.~Frittelli and O.~A. Reula.
\newblock First-order symmetric hyperbolic {E}instein equations with arbitrary
  fixed gauge.
\newblock {\em Phys. Rev. Lett.}, 76:4667--4670, 1996.

\bibitem{BONA-97}
C.~Bona and J.~Masso.
\newblock First order hyperbolic formalism for numerical relativity.
\newblock {\em Phys. Rev.}, D 56:1022--1026, 1997.

\bibitem{BONA-97-1}
C.~Bona, J.~Masso, E.~Seidel, and J.~Stela.
\newblock First order hyperbolic formalism for numerical relativity.
\newblock {\em Phys. Rev.}, D 56:3405--3415, 1997.

\bibitem{ANDERSON-98}
A.~Anderson and Jr. J.~W.~York.
\newblock Hamiltonian time evolution for general relativity.
\newblock {\em Phys. Rev. Lett.}, 81:1154, 1998.

\bibitem{ARBONA-99}
A.~Arbona, C.~Bona, J.~Masso, and J.~Stela.
\newblock Robust evolution system for numerical relativity.
\newblock {\em Phys. Rev.}, D 60,:104014, 1999.

\bibitem{ALCUBIERRE-99}
M.~Alcubierre, B.~Brugmann, M.~Miller, and Wai-Mo Suen.
\newblock Conformal hyperbolic formulation of the {E}instein equations.
\newblock {\em Phys. Rev.}, D 60:064017, 1999.

\bibitem{BAUMGARTE-99}
T.~W. Baumgarte and S.~L. Shapiro.
\newblock On the numerical integration of {E}instein's field equations.
\newblock {\em Phys. Rev.}, D59:024007, 1999.

\bibitem{BRODBECK-99}
O.~Brodbeck, S.~Frittelli, P.~Hubner, and O.~A. Reula.
\newblock The {C}auchy problem and the initial boundary value problem in
  numerical relativity.
\newblock {\em Journal of Math Physics}, 40:909, 1999.

\bibitem{LAGUNA-99}
P.~Laguna.
\newblock Linear-nonlinear formulation of {E}instein equations for the two-body
  problem in general relativity.
\newblock {\em Phys. Rev.}, D 60:084012, 1999.

\bibitem{SCHEEL-99}
M.~A. Scheel, T.~W. Baumgarte, G.~B. Cook, and S.~L. Shapiro.
\newblock Treating instabilities in a hyperbolic formulation of {E}instein's
  equations.
\newblock {\em Phys. Rev.}, D58:024007, 1999.

\bibitem{SHIBATA-99}
M.~Shibata.
\newblock Fully general relativistic simulation of merging binary clusters -
  spatial gauge condition.
\newblock {\em Prog. Theor. Phys.}, 101:1199--1233, 1999.

\bibitem{SHIBATA-99-2}
M.~Shibata.
\newblock Fully general relativistic simulation of coalescing binary neutron
  stars: Preparatory tests.
\newblock {\em Phys. Rev.}, D60:104052, 1999.

\bibitem{ALCUBIERRE-00-1}
M.~Alcubierre, B.~Brugmann, T.~Dramlitsch, J.~A. Font, P.~Papadopoulos,
  E.~Seidel, N.~Stergioulas, and R.~Takahashi.
\newblock Towards a stable numerical evolution of strongly gravitating systems
  in general relativity: {T}he conformal treatments.
\newblock {\em Phys. Rev.}, D 62:044034, 2000.

\bibitem{SHINKAI-00}
Hisa{-}aki Shinkai and Gen Yoneda.
\newblock Hyperbolic formulations and numerical relativity: experiments using
  {A}shtekar's connection variables.
\newblock {\em Class. Quantum Grav.}, 17:4799--4822, 2000.

\bibitem{KELLY-01}
B.~Kelly, P.~Laguna, K.~Lockitch, J.~Pullin, E~Schnetter, D.~Shoemaker, and
  M.~Tiglo.
\newblock Cure for unstable numerical evolutions of single black holes:
  {A}djusting the standard {ADM} equations in the spherically symmetric case.
\newblock {\em Phys. Rev.}, D 64:084013, 2001.

\bibitem{KIDDER-01}
L.E. Kidder, M.A. Scheel, and S.A. Teukolsky.
\newblock Extending the lifetime of 3{D} black hole computations with a new
  hyperbolic system of evolution equations.
\newblock {\em Phys. Rev.}, D 64:064017, 2001.

\bibitem{ANNINOS-95-2}
P.~Anninos, K.~Camarda, J.~Masso, E.~Seidel, W.~Suen, and J.~Towns.
\newblock Three-dimensional numerical relativity: {T}he evolution of black
  holes.
\newblock {\em Phys. Rev.}, D 52:2044--2058, 1995.

\bibitem{BONA-98}
C.~Bona, J.~Masso, E.~Seidel, and P.~Walker.
\newblock Three dimensional numerical relativity with a hyperbolic formulation.
\newblock 1998.
\newblock gr-qc/9804052.

\bibitem{SEIDEL-98}
E.~Seidel.
\newblock Numerical relativity: Towards simulations of 3{D} black hole
  coalescence.
\newblock 1998.
\newblock gr-qc/9806088.

\bibitem{SEIDEL-99}
E.~Seidel and W.-M. Suen.
\newblock Numerical relativity as a tool for computational astrophysics.
\newblock {\em Journal of Computational and Applied Mathematics}, 109:493--525,
  1999.

\bibitem{ALCUBIERRE-01-2}
M.~Alcubierre, W.~Benger, B.~Bruegmann, G.~Lanfermann, L.~Nerger, E.~Seidel,
  and R.~Takahashi.
\newblock The 3{D} grazing collision of two black holes.
\newblock 2001.
\newblock gr-qc/0012079.

\bibitem{MILLER-00}
M.~Miller.
\newblock On the numerical stability of the {E}instein equations.
\newblock 2000.
\newblock gr-qc/0008017.

\bibitem{TEUKOLSKY-00}
S.~A. Teukolsky.
\newblock Stability of the iterated {C}rank-{N}icholson method in numerical
  relativity.
\newblock {\em Phys. Rev.}, D 61:087501, 2000.

\bibitem{CH-II}
R.~Courant and D.~Hilbert.
\newblock {\em Methods of Mathematica Physics}, volume~II.
\newblock John Wiley \& Sons, New York, 1989.

\bibitem{WILLIAMS-80}
W.~E. Williams.
\newblock {\em Partial Eifferential Equations}.
\newblock Clarendon Press, Oxford, 1980.

\bibitem{FISHER-72}
A.~Fischer and J.~Marsden.
\newblock {\em Comm. Math. Phys.}, 28:1, 1972.

\bibitem{WALD}
R.~M. Wald.
\newblock {\em General relativity}.
\newblock The University of Chicago Press, Chicago, 1984.

\bibitem{LL}
L.~D. Landau and E.~M. Lifshits.
\newblock {\em The Classical Theory of Fields}.
\newblock Butterworth-Heinemann, Oxford, 1975.

\bibitem{LIFSHITS-61}
E.M. Lifshitz, V.V. Sudakov, and I.M. Khalatnikov.
\newblock {\em Soviet Physics JETP}, 13:1298, 1961.

\bibitem{ALCUBIERRE-00-2}
M.~Alcubierre, G.~Allen, and B.~Brugmann.
\newblock Towards an understanding of the stability properties of the 3+1
  evolution equations.
\newblock {\em Phys. Rev.}, D 62:124011, 2000.

\bibitem{ALCUBIERRE-98}
M.~Alcubierre and J.~Masso.
\newblock Pathologies of hyperbolic gauges in general relativity and other
  field theories.
\newblock {\em Phys. Rev.}, D 57: R4511, 1998.

\bibitem{MTW}
C.~W. Misner, K.~S. Thorne, and J.~A. Wheeler.
\newblock {\em Gravitation}.
\newblock W. H. Freeman and Company, New York, 1973.

\bibitem{SMARR-77-1}
L.~Smarr and W.~York.
\newblock Radiation gauge in general relativity.
\newblock {\em Phys. Rev.}, D 17:1945--1956, 1977.

\bibitem{SMARR-77-2}
L.~Smarr and W.~York.
\newblock Kinematical conditions in the construction of spacetime.
\newblock {\em Phys. Rev.}, D 17:2529--2551, 1977.

\bibitem{RINDLER-66}
W.~Rindler.
\newblock {\em Am. J. of Phys.}, 34:1174, 1966.

\bibitem{FN}
V.~P. Frolov and I.~D. Novikov.
\newblock {\em Black Hole Physics}.
\newblock Kluwer Academic Publishers, Dordrecht, 1998.

\bibitem{ZELMANOV-56}
A.~L. Zelmanov.
\newblock {\em Docl. Acad. Nauk USSR}, 107:815, 1956.

\end{thebibliography}
\end{document}